\documentclass[%
 reprint,
 amsmath,amssymb,
 aps,
]{revtex4-1}

\usepackage{graphicx}% Include figure files
\usepackage{dcolumn}% Align table columns on decimal point
\usepackage{bm}% bold math
\usepackage{color}
\usepackage{float}
\usepackage{xcolor}
\begin{document}

%\preprint{APS/123-QED}

\title{Collisional Model of Energy Dissipation in 3D Granular Impact}% Force line breaks with \\

\author{Cacey Stevens Bester}
\email{cacey.stevens@phy.duke.edu}

\author{Robert P. Behringer}
\affiliation{%
 Department of Physics, Duke University, Durham, NC 27708
}

\date{\today}

\begin{abstract}
We study the dynamic process occurring when a granular assembly is displaced by a solid impactor.
The momentum transfer from the impactor to the target is shown to occur through sporadic, normal collisions of high force carrying grains at the intruder surface.
We therefore describe the stopping force of the impact through a collisional based model.
To verify the model in impact experiments, we determine the forces acting on an intruder decelerating through a dense granular medium using high-speed imaging of its trajectory.
By varying the intruder shape and granular target, intruder-grain interactions are inferred from the consequent path.
As a result, we connect the drag to the effect of intruder shape and grain density based on a proposed collisional model.

\begin{description}
\item[PACS numbers]
%You may use the \texttt{description} environment to structure your abstract; use the optional argument of the \verb+\item+ command to give the category of each item. 
\end{description}
\end{abstract}

\pacs{Valid PACS appear here}
                            
                              
\maketitle

\section{\label{sec:level1}Introduction}

The seemingly simple occurrence of a solid object impacting a bed of sand remarkably exposes the truly complex dual nature of granular media.
Despite the upward splash of displaced grains, the response of the granular target is to cause the intruding object to abruptly stop as its momentum is carried away through the material.
There has been significant effort to understand the way by which momentum is transferred to granular media, beginning over a century ago \cite{poncelet}.
Studies of penetration into dry granular media have expanded significantly to focus on crater formation \cite{uehara,newhall, ambroso, tsimring, seguin, umbanhowar2010,thoroddsen} and proposed governing force laws \cite{KatDur, katpre2013, nordstrom, altshuler, tiwari}.
Additionally, a complete comprehension of granular impact has a natural connection to navigation on grainy surfaces \cite{CLi,aguilar}, geomorphology, and astrophysical craters \cite{daniels, guttler, dowling, ruiz}.
However, even with such stimulating work, the grain-scale details of force transmission during granular impact have not been fully resolved \cite{KatBk}.

An empirical force law has been established to broadly describe this process.
It represents the stopping force due to granular media as the sum of depth-dependent static force $f(z)$ and velocity-dependent inertial drag $h(z)\dot{z}^2$, such that the force acting on the intruder is given by

\begin{equation}
F=m\ddot{z}=mg-f(z)-h(z)\dot{z}^2
\end{equation}

\noindent
where $mg$ is gravitational force ($m$ is the intruder mass and $g$= 9.8 $m/s^2$), $\dot{z}$ is the intruder velocity, and $h(z)$ gives the inertial drag coefficient \cite{KatDur}.
These terms highlight the dual nature of granular media.
The intruder motion is used to determine the form of the stopping force law, where the dynamics are governed by inertial drag over much of the deceleration.
Nonetheless there lacks a connection of the experimental relation to its physical origin, which would resolve remaining controversy to explain the inertial drag \cite{ruiz, KatBk}.

Clark \textit{et. al.} \cite{ClarkPRL} notably explored granular impact via two-dimensional (2D) experiments using photoelastic grains as the target, whereby the granular response could be visualized.
They showed that the momentum is lost from an intruding disk through rapidly fluctuating collisions at the disk surface with clusters of high force-carrying grains.
The clusters form chain-like pathways that carry away momentum.
The observations are connected to the inertial drag of the stopping force by way of a collisional model \cite{takehara,ClarkPRE}, a novel microscopic explanation of 2D granular impact.

Here we extend the collisional model to three-dimensional granular impact occurrences, such as the aforementioned case of a solid object impacting sand, to show this as a universal explanation of the deceleration.
We determine the forces acting on an intruder penetrating a dry, dense granular medium using high-speed imaging of its trajectory.
We relate the force to the proposition that momentum transfer occurs through rapid, sporadic collisions that act perpendicular to the impactor surface; friction is not important for this process.
Accordingly we present a model based on collisions of the intruder with granular chains during impact, and we show that this model effectively explains the stopping force of granular media on an intruder.
We vary grain density and the shape of the intruding object to infer intruder-grain interactions.
As a result, we connect the inertial drag  $h(z)\dot{z}^2$ to the effect of the intruder shape and target density based on the proposed collisional model.  

\section{\label{sec:level1}Experiment Details}

Figure 1(a) illustrates the details of the experimental setup.
In all experiments, a dry noncohesive granular medium, sand (grain density $\rho_g$ = 2.1 g/cm$^3$ and grain diameter $D_g\sim$ 0.3mm) or couscous ($\rho_g$ = 1.2 g/cm$^3$ and $D_g\sim$ 2mm), is used as the impact target.
The target is held within a large rectangular container (38 cm width by 49 cm length by 20 cm depth); this size is large enough to minimize boundary effects to the intruder's trajectory \cite{nelson, seguin}.
The packing fraction, fraction of container space that is filled by grains, is held approximately constant, as acquired by pouring grains into the container and tapping.
Before each run, we prepare the granular bed by leveling the impact surface using a straight edge.

\begin{figure}[!]
\centering
\includegraphics[width=8 cm]{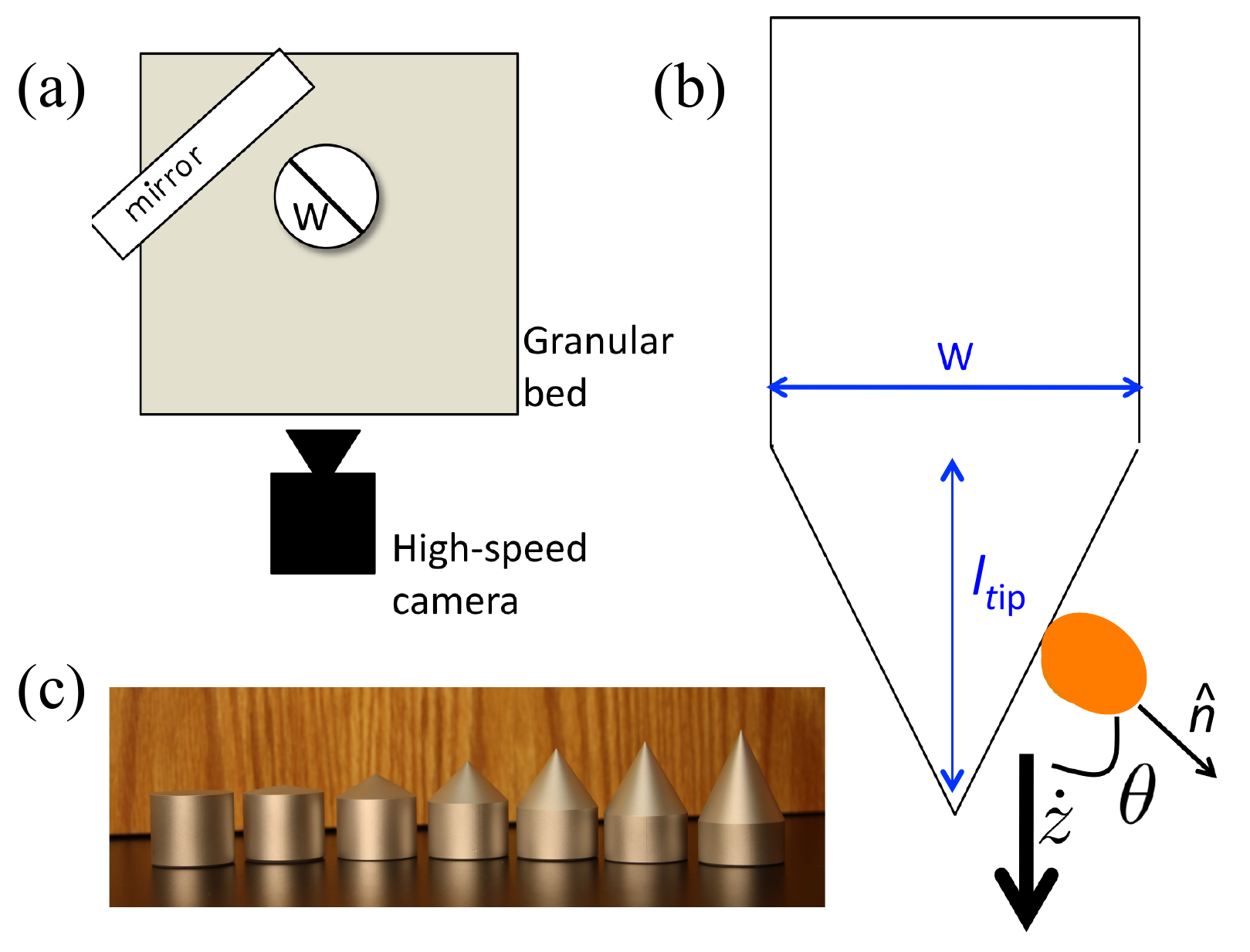}
	\caption{(a) Schematic of the top view of experiment setup. A high-speed camera is used to view an intruder which is released from a height above and impacts into the granular bed. A mirror is placed at 45 degrees to the impact to measure any tilting during penetration. (b) Diagram of the impact of a conical intruder with a chain of particles of the granular target (represented by an orange ellipse). (c) Image of all intruders of $w$=3.8cm, varying $s$ through the following values: 0, 0.2, 0.5, 1, 1.4, 1.7, 2.1.}
	\label{expt}
\end{figure}

We work with intruders of conical shapes to methodically vary the interaction between the angle of the intruder's surface with the granular target during penetration, as depicted in fig. 1(b).
The slope of the cone $s$ is the ratio of the length of the cone $l_{tip}$ to half of its width $w$ and ranges in these experiments from 0 (a cylinder) to 2.1 (a sharp cone).
We also explore the effect of the cone width $w$ using $s$=1 cones of $w$=2.2 cm to $w$=3.8 cm.
The mass is held constant at 101 g for all intruders.

An electromagnet serves as the release mechanism of the intruder, which then falls under gravity and impacts at the center of the surface of the granular bed.
It is released from a height of 6 cm to 2 m, as measured from the tip of the intruder, to achieve an impact velocity $\dot{z}_i$ extending from 1 m/s to 6 m/s.

Upon contact, the intruding object comes to rest in less than 0.1 seconds.
Accordingly, videos of impact runs are captured at 30,000 frames per second. 
We use a Photron Fastcam SA5 high speed camera that is placed at a side view to the granular bed.
We use these images to determine the position $z$ during the time $t$ of its trajectory, where one pixel of resolution corresponds to 0.04 cm.
A thin rod is attached to each intruder to track $z$ during the penetration.
We also place a mirror at 45 degrees to the impact point to measure tilting of the intruder. 
To obtain velocity $\dot{z}$, we take the derivative of $z(t)$ and reduce the amplified noise of the data by performing a convolution with a gaussian filter of width of 2.5\% of the length of $z(t)$ for each run \cite{conv}.
The moment of impact is then identified by imaging in conjunction with locating the maximum peak of the $\dot{z}(t)$ curve.
We differentiate once more to find the intruder acceleration $\ddot{z}(t)$; the noise of $\ddot{z}$ is similarly filtered via convolution with a gaussian filter.

\begin{figure}[!]
\label{trajectory}
\centering
\includegraphics[width=7 cm]{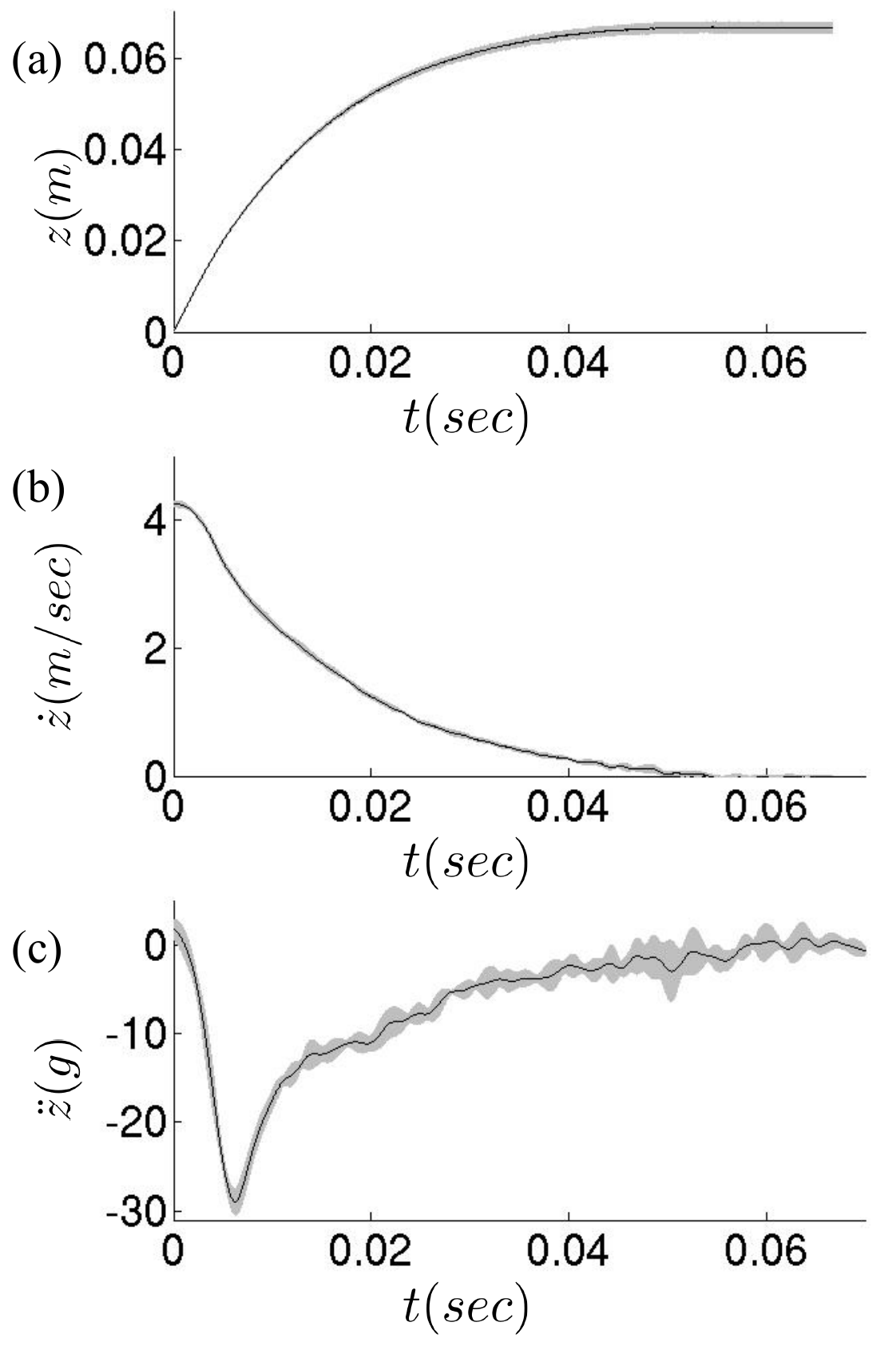}
\caption{(a) $z$ vs. $t$, (b) $\dot{z}$ vs. $t$, and (c) $\ddot{z}$ vs. $t$ of a $s$=1 intruder impacting couscous. The gray region surrounding each curve indicates the standard deviation of five runs.} 
\end{figure}

\section{\label{sec:level1}Tracking Trajectories}

\subsection{\label{sec:level2} Intruder Dynamics}

As an object impacts and penetrates the dense granular bed, it experiences a strong drag force which causes an abrupt stop.
The dynamics of this process are shown in fig. 2 as the mean of five runs of an $s$ =1 intruder penetrating couscous; the gray shaded regions of each curve indicate the standard deviation of the five measurements.
As given in fig. 2(a), $z$ rapidly approaches and later saturates at a maximum penetration depth $z_{stop}$.
Note that $z$ is positive and increasing as the intruder moves deeper in the granular target, with $z$=0 and $t$=0 at impact.
We then differentiate $z(t)$ to find $\dot{z}(t)$ of fig. 2(b);
$\dot{z}(t)$ declines to zero within 0.05 seconds.
From $\dot{z}(t)$, we determine the stopping time $t_{stop}$ as the time from initial contact to the point at which $\dot{z}(t)$ first reaches 0.
In fig. 2(c) we show $\ddot{z}(t)$, which is found by differentiating $\dot{z}(t)$; $\ddot{z}(t)$ reflects forces exerted on the intruder by the granular target.
There is a strong peak within milliseconds of impact, followed by a more gradual deceleration to zero.
The fluctuations in $\ddot{z}(t)$ are clear and due to the intermittent emission of energy.
The details of the observed fluctuations are influenced by our imaging sample rate and noise filtering approach.

The $z(t)$ and $\dot{z}(t)$ trajectories are reproducible within small variations.
However, $\ddot{z}(t)$ has more amplified noise in part due to repeated differentiation and convolution which can distort the curves.
It is therefore difficult to resolve force fluctuations from the granular medium, which have been 
observed in other experiments as an indication of the building and breaking up of force chains \cite{ClarkPRL, altshuler, goldman}.
We compare $\ddot{z}(t)$ determined from high-speed imaging to that of direct force measurements via an accelerometer (Analog Devices ADXL377, sample rate = 500 Hz), where the imaging and accelerometer measurements are taken simultaneously (see fig. 3).
The overall shape of the $\ddot{z}(t)$ curve matches when determined by both methods, yet exact fluctuations are difficult to resolve due to noise filtering effects.
The gaussian filter width smoothes the curve, while it can also cause features of data to be distorted \cite{conv}.
Accordingly, an accelerometer with a higher sample rate is needed to address how acceleration fluctuations change with intruder properties, and this will be the subject of future work.

\begin{figure}[h]
\label{trajectory}
\centering
\includegraphics[width=7 cm]{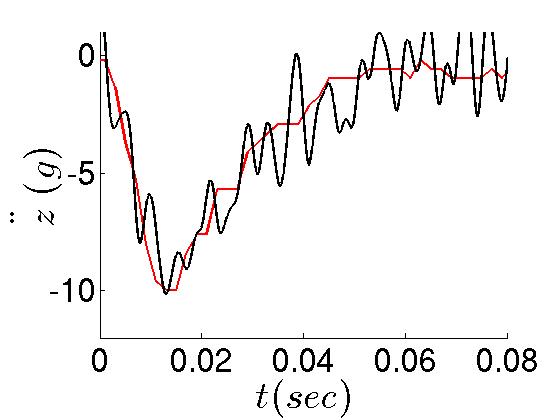}
\caption{Comparison of $\ddot{z}(t)$ of $s$=2.1 intruder impacting couscous as determined simultaneously by high speed imaging (black line) and by direct force measurements via accelerometer (red line). The shape of the curves matches for these approaches, though force fluctuations cannot be resolved well. } 
\end{figure}

\subsection{\label{sec:level2} Intruder Shape Effect}

\begin{figure*}[!]
\label{PVs}
\centering
\includegraphics[width=13 cm]{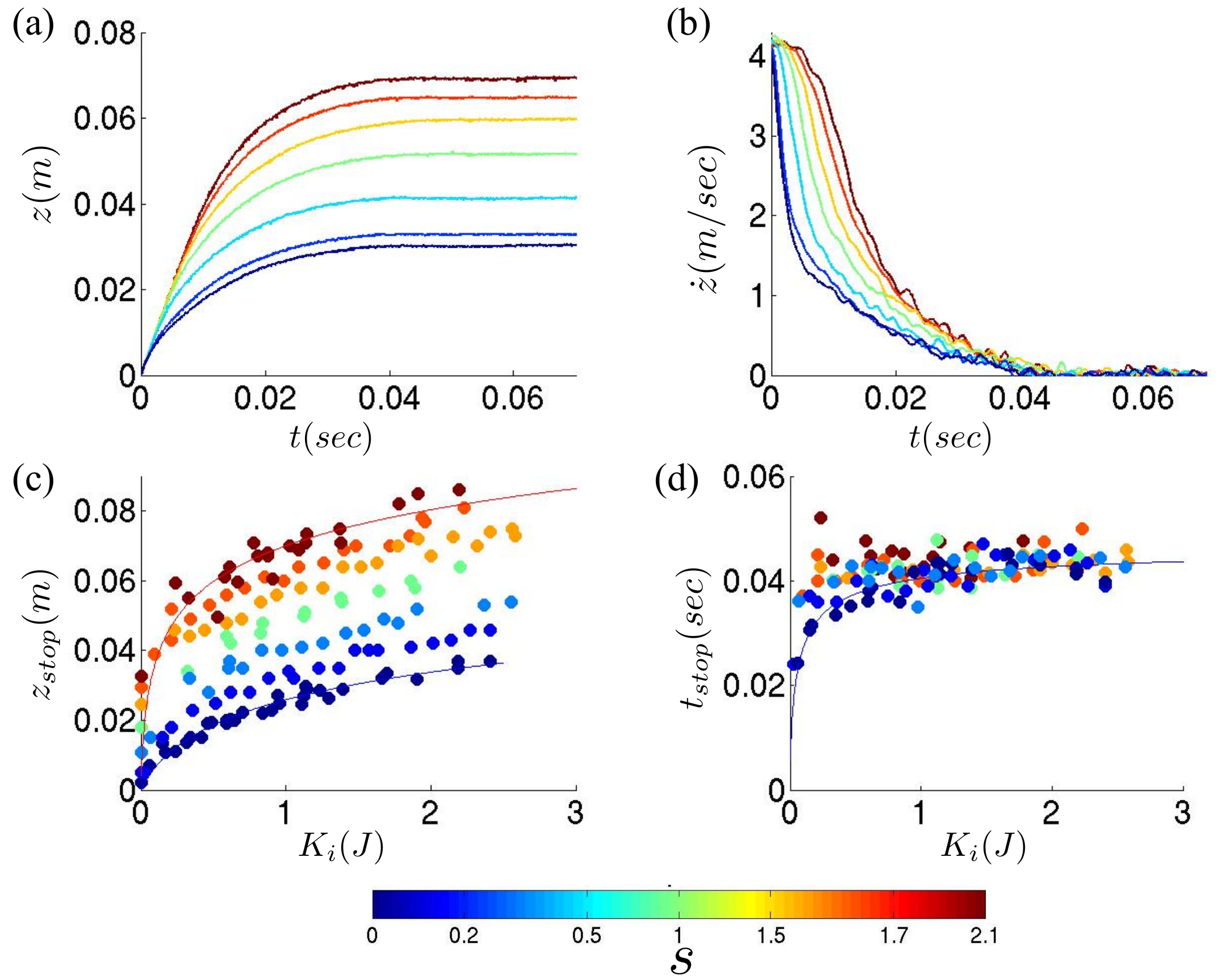}
\caption{(a) $z$ vs. $t$, varying $s$, all impacting sand from the same release height. Penetration depth $z_{stop}$ increases with increasing $s$. 
(b) $\dot{z}$ vs. $t$. The decline of $\dot{z}(t)$ changes with increasing $s$, but all curves reach $\dot{z}$=0 at approximately the same time.
(c) $z_{stop}$ vs. $K_i$. Higher $s$ shifts curve to higher values but does not change the trend. The fit of $z_{stop}\propto a\ log (dK_i+1)$ where $a$ and $d$ are constants, is shown with the data of the $s$=0 and $s$=2.1 intruders.
(d) $t_{stop}$ vs. $K_i$. Above 1 J, $t_{stop}$ is constant for all $K_i$ and $s$.
Solid line shows the fit of eq. 4 to data $t_{stop}$($K_i$) of $s$=0 intruder.
Data sets correspond to $s$ values given by the colorbar. } 
\end{figure*}

Figure 4 gives the effect of cone height, as defined by $s$, on the intruder penetration dynamics.
We start by observing $z(t)$ at constant release height (constant initial velocity $\dot{z}_i$).
When we vary $s$, we find similar trajectories of $z(t)$ as the intruder rapidly approaches $z_{stop}$.
However different $z_{stop}$ values are achieved; $z_{stop}$ increases with $s$.
Specifically, sharper cones achieve deeper penetration into the granular target, given the same initial energy.
This behavior extends to the change of $\dot{z}(t)$ curves with $s$, as shown in fig. 4(b): $\dot{z}$ declines less rapidly with increasing $t$ as we increase $s$, though $\dot{z}$ reaches 0 at approximately the same $t_{stop}$.

We extend the behavior of $z_{stop}$ and $t_{stop}$ to describe their dependence on initial kinetic energy $K_i=1/2m\dot{z_i}^2$ and connect to our study of the empirical drag force law (eq.1).
Previous authors have shown that eq. 1 can be solved by reframing to the kinetic energy $K$ versus $z$ form where $K=1/2m\dot{z}^2$ and $m\ddot{z}=dK/dz$ \cite{ambroso, clarkepl}.
This formulation leads to the following:

\begin{equation}
\frac{dK}{dz}=mg-f(z)-\frac{2h(z)}{m}K
\end{equation}

\noindent
To obtain expressions for $z_{stop}$ and $t_{stop}$ from eq. 2, forms of $f(z)$ and $h(z)$ are typically assumed to be constant \cite{tsimring}.
If we take $f(z)$=$f$ and $h(z)$=$b$, we first attain an equation for $z_{stop}(K_i)$ as 

\begin{equation}
z_{stop}=\frac{m}{2b}ln \left[1+\frac{2b}{m(f-mg)}K_i \right]
\end{equation}

\noindent
Note that this expression is no longer relevant at $K_i$=0 since $z_{stop}$($K_i$=0)$>$0 for any intruder.
Figure 4(c) shows the dependence of $z_{stop}$ on $K_i$.
As similarly shown with previous experiments of 2D impact \cite{clarkepl}, $z_{stop}$ logarithmically rises with $K_i$.
The fit of eq. 3 is shown in fig. 4(c) for an $s$=0 intruder.
It successfully captures the trend of $z_{stop}$($K_i$).

We systematically vary $s$ to determine how $z_{stop}$ changes with cone shape.
The curve shifts to higher values with increasing $s$, though the simple fit (eq. 3) does not apply well for $s>$0, as displayed, for example, for $s$=2.1 intruder.
We propose that this is an indication that the inertial drag coefficient $h(z)$ is not constant for sharper cones,
implying a connection between intruder shape and inertial drag.

The dependence of $t_{stop}$ on $K_i$ can also be determined from eq. 1 using constant $h(z)$ and $f(z)$ \cite{goldman} such that

\begin{equation}
t_{stop}=\frac{tan^{-1}[(\frac{2bK_i}{m(f-mg)})^\frac{1}{2}]}{[b(\frac{f}{m^2}-\frac{g}{m})]^\frac{1}{2}}
\end{equation}

\noindent
The plot of $t_{stop}$ versus $K_i$ is given with the fit of eq. 4 in fig. 4(d).
We find an approximately constant $t_{stop}$ is consistent for high $K_i$.
Above $K_i\sim$0.5 J, $t_{stop}$ reaches 0.04$\pm$0.005 sec. for all $s$.
Again, we find that the solution (eq. 4) is a reasonable fit to the experimental data of $s$=0.
For $s>$0, eq. 4 only corresponds with the trend of data for high $K_i$.
However, for $s$=0 intruders, these scalings are also expected to depend on the granular target, as we discuss later in this paper.

\section{\label{sec:level1}Collisional Model}

\subsection{\label{sec:level2} Description}

We use the dependence of the impact trajectory on intruder shape to demonstrate that normal collisions capture inertial drag.
If deceleration is achieved through collisions acting perpendicular to the surface, a change in intruder shape would predictably result in different inertial drag ($h(z)\dot{z}^2$) values. 
Accordingly, we attain an expression in terms of the shape of the intruder to define $h(z)\dot{z}^2$.
Note that this approach is similar to that of 2D impact experiments with photoelastic grains \cite{ClarkPRE}.

Figure 1(b) shows the schematic of a cone shaped intruder colliding with a cluster of the granular target, where the orange ellipse represents the chain of particles impacted by the surface and $\theta$ is the angle between intruder velocity $v$ and chain of particles at surface normal $n$.
Force is then perpendicular to the intruder surface.
The force due to this intruder-grain collision can be expressed as the ratio of change in momentum $\Delta p \propto m_gvcos\ \theta$ to time $t=\Delta d/(v\ cos\ \theta)$ such that

\begin{equation}
f=\frac{\Delta p}{\Delta t}\propto \frac{m_gv^2cos^2 \theta}{d}
\end{equation}

\noindent
where $d$ is the grain diameter and $m_g$ is the grain mass.
The number of discrete collisions possible across the surface is given by $dS/d^2$.
The total force is then calculated by integrating over the surface area:

\begin{equation}
F=\int f \frac{dS}{d^2} \propto \int \frac{m_gV^2 cos^2\theta   dS}{d^3}
\end{equation} 

The variables $dS$ and $cos\ \theta$ give the shape effect to the drag force.
The width of the intruder is expressed through $dS$, whereas $cos\ \theta$ can be written in terms of the slope of the cone $s$ as

\begin{equation}
cos\ \theta=\frac{dx}{dl}=\frac{dx}{\sqrt{dx^2 + dz^2}}=\frac{1}{\sqrt{1+s^2}}
\end{equation}

\noindent
where $s$=$dz/dx$.
Finally, this provides a description of the inertial drag based on the collisional model for all possible collisions over the intruder surface,

\begin{equation}
h(z)\dot{z}^2 \propto \frac{\rho_g \dot{z}^2 w^2}{(1+s^2)}
\end{equation}

\noindent
where $\rho_g$ is the grain density and $w$ is the width of the portion of the cone that is penetrated in the granular media. 
We thereby obtain a simple scaling relation for the inertial drag force term that is quadratic with velocity and depends on intruder shape and grain density.

\begin{figure}[!]
\label{trajectory}
\centering
\includegraphics[width=6 cm]{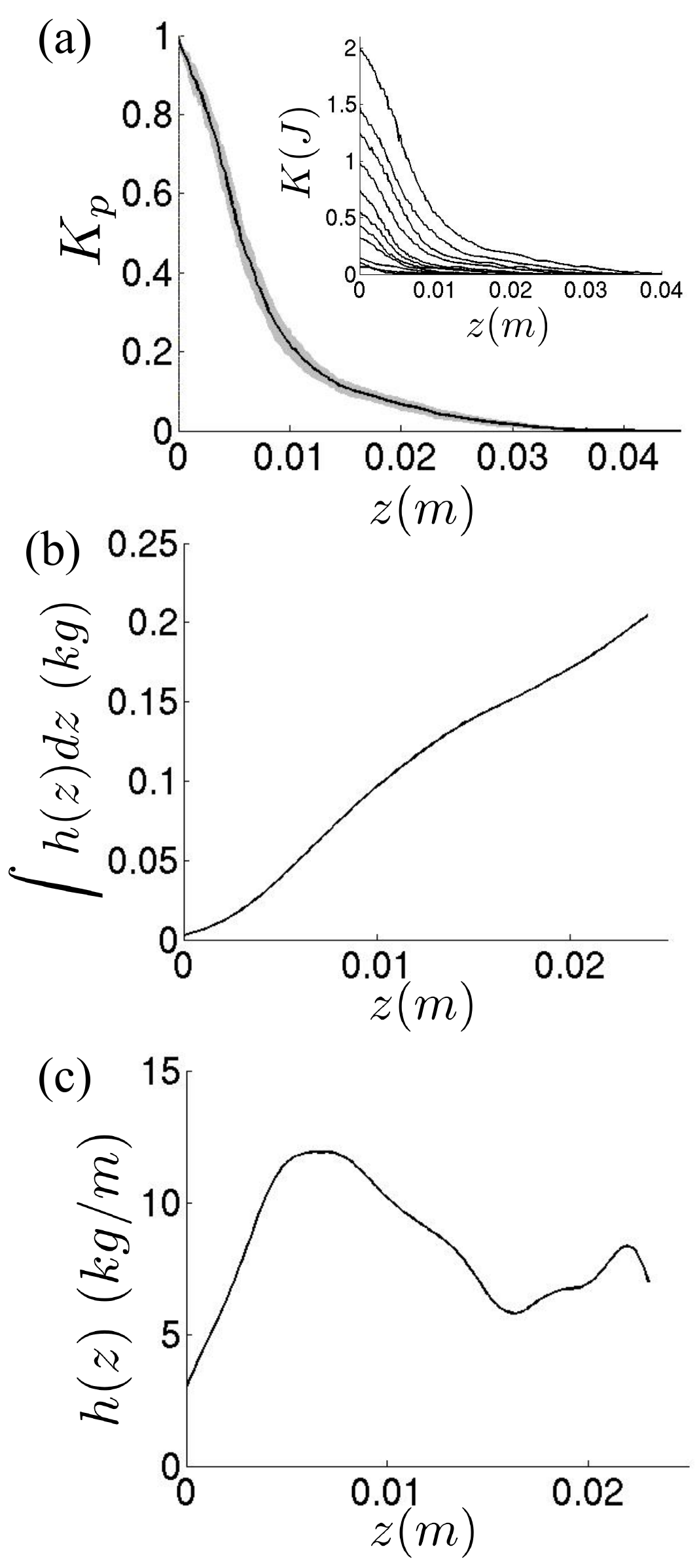}
\caption{(a) Inset: $K$ versus $z$ of $s$=0 intruder impacting sand. Each curve represents a trajectory of a different impact velocity. Main panel: $K_{p}$ versus $z$, as determined from all pairs of $K(z)$ trajectories of the inset plot. The gray region surrounding the curve indicates the standard deviation from the mean of $K_p$. (b) $\int h(z)dz$ versus $z$ (c) $h(z)$ versus $z$ for a $s$=0 intruder impacting sand, showing oscillations about a constant value.} 
\end{figure}

We now demonstrate that the collisional model expression gives the grain-scale origin of inertial drag of 3D granular impact, $h(z)\dot{z}^2$, which we determine experimentally.
It could be found via acceleration data, which we have shown can present significant error.
Additionally, using the force law to find $h(z)$ would require determining $z$, $\dot{z}$, and $\ddot{z}$ for many trajectories of each intruder.
We therefore apply eq. 2 where the force law model is reformulated into a differential equation in kinetic energy \citep{clarkepl}.
We acquire $h(z)$ from $K$ versus $z$ so that we can compare the result to the proposed collisional model.
Without any assumptions of the form of $h(z)$ and $f(z)$, the differential equation is solved by means of the integrating factor method using $e^{\int \frac{2h(z)}{m}dz}$ to achieve  

\begin{equation}
K(z)=K_p(K_i + \phi)
\end{equation}

\noindent
where $K_p=e^{-\int \frac{2h(z)}{m} dz}$, and $\phi = \int (mg-f(z))e^{\int \frac{2h(z)}{m} dz} dz$.  

We can take the difference of the differential equation of two $K(z)$ curves to isolate $h(z)$.
We thereby attain the relation

\begin{equation}
K_p= \frac{K_b-K_a}{K_{b,i}-K_{a,i}}=exp^{-\int \frac{2h(z)}{m} dz}
\end{equation}

\noindent
where $a$ and $b$ indicate indices of different trajectories.
We then determine $h(z)$ for each intruder by comparing all pairs of $K(z)$ trajectories.

Figure 5 shows how we determine $h(z)$ for an $s$=0 intruder penetrating sand.
We start by expressing data as $K(z)$ at varying $\dot{z}_i$ (shown in the inset of fig. 5(a)) to use the kinetic energy approach. 
There is a rapid decline of $K(z)$ during penetration.
The average of differences of all pairs of $K(z)$ trajectories is then determined (see fig. 5(a)).
We use $K_p$ to solve for $h(z)$ as follows,
\[\int h(z)dz=-\frac{m}{2}ln \ K_{p}\]

This is given in fig. 5(b). 
The result in this case is a $\int h(z) dz$ curve that is an approximately linear function of $z$.
We use a cutoff of $K_p<$0.05 since the static term $f(z)$ becomes dominant as the intruder comes to rest.
The derivative of this curve gives $h(z)$ and requires noise reduction by performing a convolution to the data (see fig. 5(c) for result).
There are oscillations in $h(z)$ which may connect to the force fluctuations and discretization of collisions;
a precise determination of these features would require direct force measurement.
To avoid noise amplification and possible data distortion introduced by taking derivative and convolution of $\int h(z) dz$, we utilize $\int h(z) dz$ for the intruder shape and grain density comparisons.

\subsection{\label{sec:level2} Shape Effect}

\begin{figure}[!]
\centering
\includegraphics[width=7 cm]{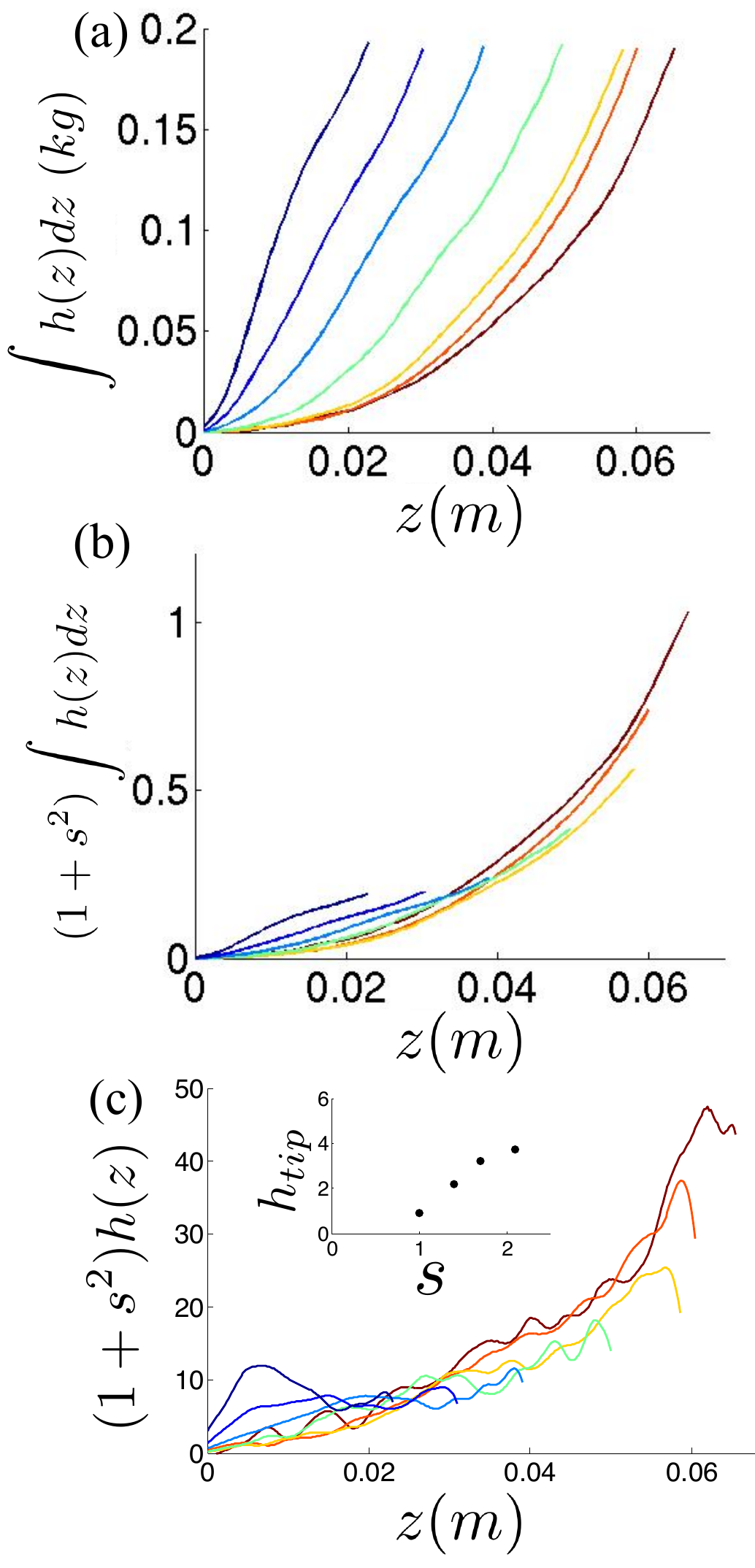}
\caption{(a) $\int h(z)dz$ versus $z$ for all intruders, ranging from $s$=0 to $s$=2.1. Colors correspond with $s$ values of fig. 4. Higher $s$ lowers the curve to smaller inertial drag values. The trend in data also changes from approximately linear to increasingly curved. (b) The data collapse to one curve when scaled as a function of $s$. (c) Collapse with consideration of added contribution from tip for $s>$1 intruders.} 
\end{figure}

Starting with the intruder shape, we show that $s$ and $w$ affect $h(z)$ as described by the collisional model expression (eq. 8).
In fig. 6(a), $\int h(z) dz$ versus $z$ is plotted as $s$ is systematically changed from 0 up to 2.1.
With increasing $s$, curves shift to lower inertial drag values.
Additionally, we find that $\int h(z) dz$ vs. $z$ plots become increasingly curved as $s$ increases.
We propose that the increased curvature results from the changing surface area at low $z$ as a sharper cone is penetrated deeper into the granular target.
Since the effective width of the cone increases from $z$=0 up to                                                                                                                                                                                                                                                                                                                                                                                                                                                                                                                                                                                                                                                                                                                                                                                                                                                                                                                                                                                                                                                                                                                                                                                                                                                                                                                                                                                                                                                                                                                                                                                                                                                                                                                                                                                                                                                                                                                                                                                                                                                                                                                                                                                                                                                                                                                                                                                                                                                                        $z$=$l_{tip}$ and $h(z)$ is proportional to $w^2$, $\int h(z)dz$ is expected to rise quadratically in this range.
As $s$ increases, the intruder must penetrate deeper to reach constant $w$.
After $z>l_{tip}$ of each intruder, $\int h(z) dz$ should become a linear function of $z$.

All curves fall onto a common curve when the $\int h(z) dz$ axis is rescaled by a function of $s$, as shown in fig. 6(b).
Accordingly, the expression $1+s^2$, as determined from the collisional-based expression for $h(z)$ (eq. 8), leads to a good collapse of all data.
This scaling provides evidence that the collisional model does connect with our 3D experiments.

Clark et. al. used photoelastic observations of two-dimensional impact experiments to show the distinct contributions from the sides and pointed tips of triangular intruders to the collisional drag coefficient $h(z)$(see fig. 4 of \citep{ClarkPRE}).
There was found to be a strong contribution of stress at the tip of pointed shapes, which was considered separately.
The contribution from the tip was found to be significant for $s>$1 and stayed approximately constant;
A fit was given by $0.2(1-e^{-2s})$.

We expect that a distinct contribution from the cone tip is relevant at high $s$ in 3D experiments as well.
Accordingly we write a new expression for $s>$1 cones: $h(z)=h_{side}(s)+h_{tip}(s)$. 
For $s\leq$1, $h(z)$ is determined by eq. 8.
To find $h_{tip}$ for 3D experiments, we first calculate the slope of each $\int h(z) dz$ curve of fig. 6(b), using only data for which $z>l_{tip}$ of each cone.
The slope of $s\leq$1 curves gives $h(z)$, while the slope of $s>$1 curves includes the tip addition.
The difference of these values gives the approximate contribution of the tip for $s>$1, shown in the inset of fig. 6(c).
In fig. 6(c), we show our $h(z)$ collapse with the added tip stress.

\begin{figure}[h]
\label{width}
\centering
\includegraphics[width=7 cm]{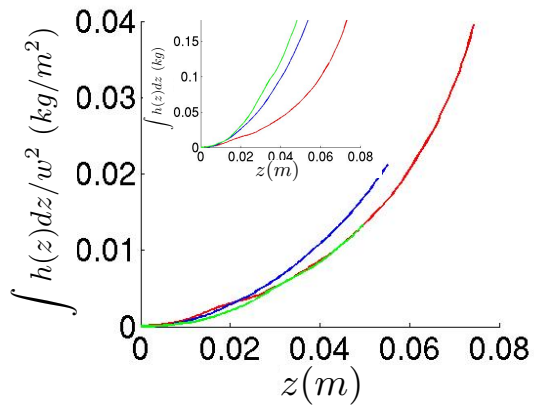}
\caption{Inset: $\int h(z)dz$ versus $z$ of $s$=1 intruder, varying width $w$: 2.2cm (red), 3.0cm (blue), and 3.8cm (green). Lower $w$ decreases the curve, but, in this case, the trend in data is not affected. Main panel: When scaled as a function of $w$, the data collapse to a single curve.} 
\end{figure}

When extending the study to the effect of $w$, we similarly observe that the collisional model applies.
Figure 7 shows $\int h(z) dz$ as a function of $z$ for $s$=1 intruders of three widths: $w$=2.2cm, $w$=3.0cm, and $w$=3.8 cm. 
The intruder mass is held constant.
The proposed shape effect is again verified by changing $w$.
The intruder penetrates deeper into the granular bed with lower $w$, and $\int h(z) dz$ decreases with decreasing $w$ as displayed in the inset plot.
The dependence of inertial drag on $w$ is then displayed by a collapse of data as $\int h(z) dz \propto w^2$.
These data rescale reasonably well to a single curve when plotted as $\int h(z) dz/w^2$, as shown in fig. 7.

\subsection{\label{sec:level2} Granular Target}

\begin{figure*}[h]
\label{grain}
\centering
\includegraphics[width=13 cm]{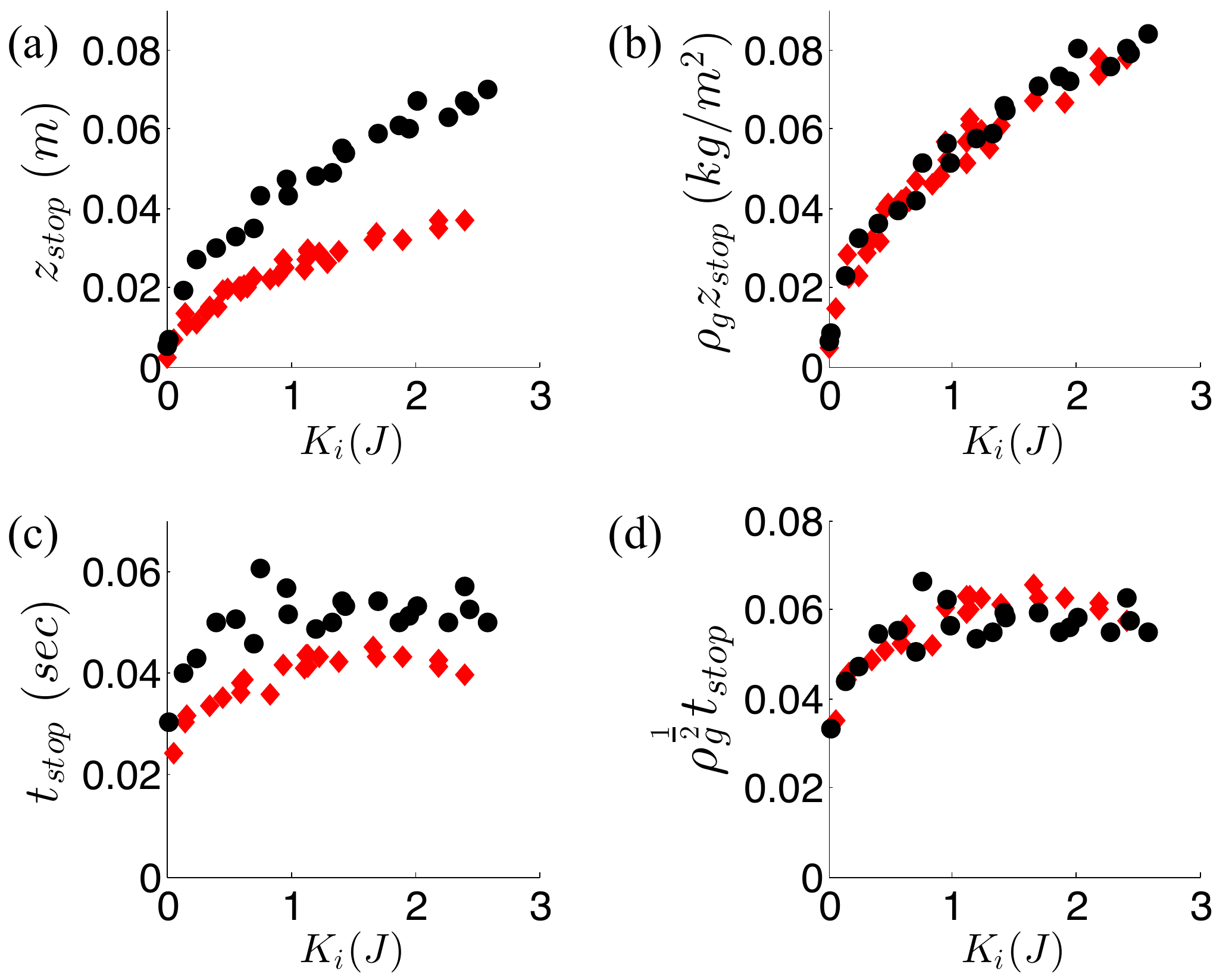}
\caption{(a) $z_{stop}$ versus $K_i$, varying target grain density $\rho_g$ to sand ($\color{red} \blacklozenge$) or couscous ($\bullet$). In each case, $z_{stop}$ shows similar dependence on $K_i$, yet the intruder penetrates to higher $z_{stop}$ with decreasing $\rho_g$. 
(b) When scaled as a function of $\rho_g$, the data collapse to a single curve.
(c) $t_{stop}$ versus $K_i$, varying $\rho_g$. Stopping time $t_{stop}$ is constant for $K_0>$0.5 J. The behavior for sand and couscous are similar; however, $t_{stop}$ increases for decreased $\rho_g$.
(d) The data collapse when scaled by $\rho_g^{1/2}$. All data is determined using an $s$=0 intruder.} 
\end{figure*}

We vary the granular target to study the effect of grain density $\rho_g$.
Figure 8(a) shows $z_{stop}$ vs. $K_i$ of a $s$=0 intruder impacting sand or couscous, where both targets are prepared by pouring grains into the container and tapping to settle the surface.
As with previous experiments (\cite{clarkepl} and fig. 4(c)), there is a gradual increase of $z_{stop}$ with increasing $K_i$, showing a logarithmic dependence.
This trend is consistent to varying $\rho_g$.
The intruder simply penetrates to higher $z_{stop}$ measurements for lower $\rho_g$.
The fit of $z_{stop}(K_i)$ (eq. 3) predicts that $z_{stop}(K_i)\propto 1/h(z)$; for the case of an $s$=0 intruder, $h(z)$ is approximately constant (defined as b in eq. 3).
By applying eq. 8, we quantify the expected dependence of $z_{stop}$ on $\rho_g$ as $z_{stop}\propto 1/\rho_g*ln K_i$.
The effect of $\rho_g$ is expressed by scaling the curves as $\rho_g*z_{stop}$, as shown in fig. 8(b).
Therefore, the successful collapse supports our proposed collisional model for the inertial drag coefficient.

For each granular target, we also measure $t_{stop}$ versus $K_i$ using an $s$=0 intruder (see fig. 8(c)).
Above $K_i>$ 0.5 J, we similarly find a constant $t_{stop}$ with increasing $K_i$,  with each granular target, as was found with changing intruder shape.
However, $t_{stop}$ increases with decreasing $\rho_g$; this reflects a slower deceleration for a low density granular target.
We again expect $t_{stop}$ to obey the scaling of eq. 4 with constant $h(z)$.
The $t_{stop}$ data collapse by $\rho_g$ as $t_{stop}~\rho_g^{1/2}$, as shown in fig. 8(d).
Accordingly, this scaling behavior with $\rho_g$ also connects to the solution of $t_{stop}$.
Therefore, we have effectively linked the experimentally determined $t_{stop}$ and $z_{stop}$ scalings with our proposed model of the drag force.

\begin{figure}[h]
\label{trajectory}
\centering
\includegraphics[width=7 cm]{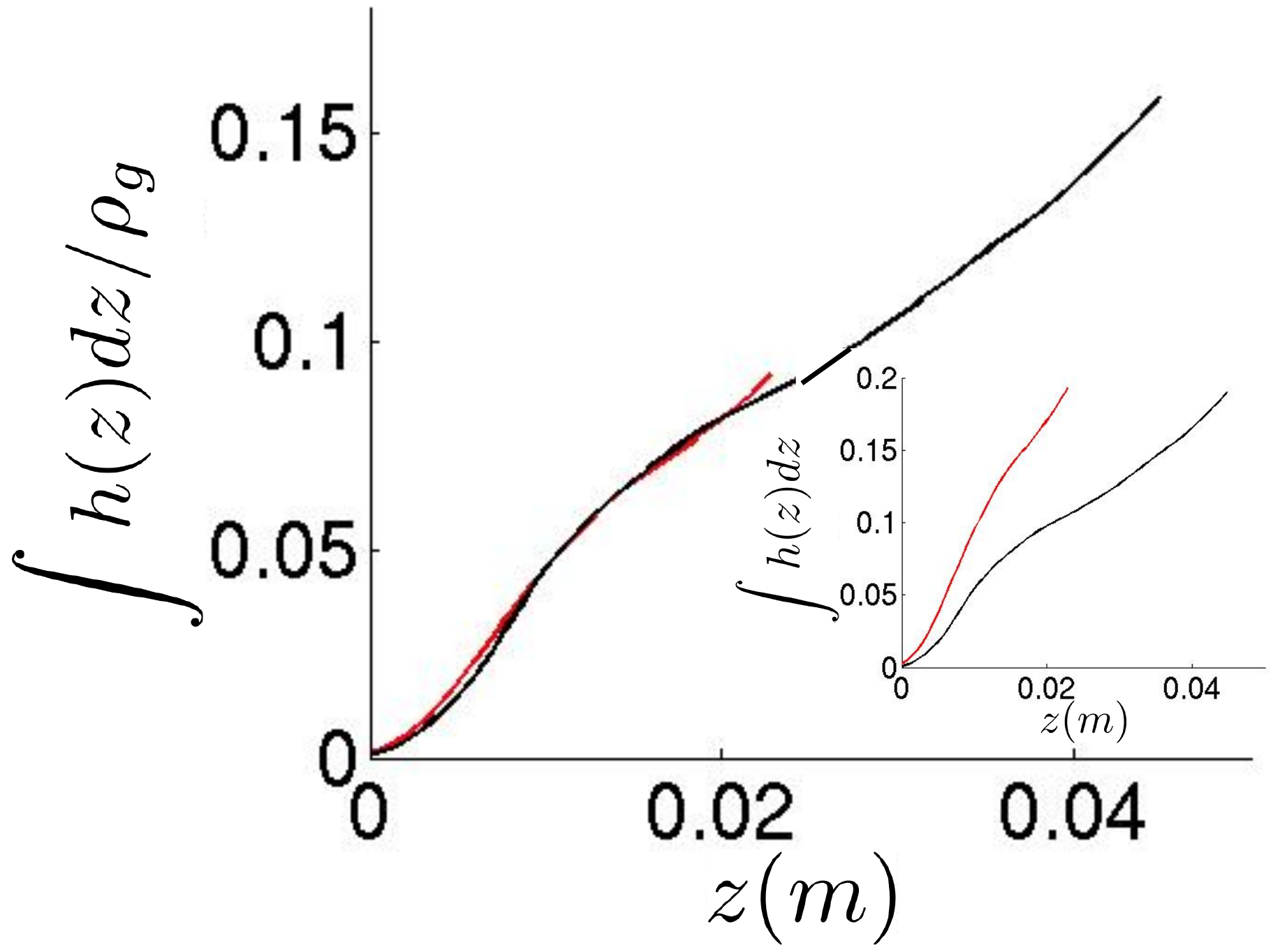}
\caption{Inset: $\int h(z)dz$ versus $z$ for $s$=0 intruder, varying $\rho_g$: sand (red line) or couscous (black line). Higher $\rho_g$ (sand) increases inertial drag $h(z)$. Main panel: Scaled $h(z)$, $\int h(z)dz/\rho_g$ versus $z$ for the two granular targets, collapsing data.} 
\end{figure}

Via $h(z)$ calculations, we further confirm that our collisional model also captures the effect of $\rho_g$ (see Fig. 9).
The inertial drag coefficient $h(z)$ is determined for an $s=0$ intruder penetrating the sand or couscous target.
For $s$=0, $\int h(z)dz$ is nearly a linear function of $z$. 
In other words, $h(z)$ does not depend on $z$ for cylindrical shaped intruders.
We find that $\int h(z) dz$ is larger for higher $\rho_g$, as expected from the $h(z)$ expression of the collisional model.
Accordingly, there is a higher stopping force due to sand, and this affects the maximum penetration depth of the intruder.
The two curves collapse very well to a single curve, when scaled by grain density effect ascribed via collisional model.

\section{\label{sec:level1}Conclusions}

We provide an extensive study of the impact of solid intruders with a dry granular target.
The deceleration of an intruder during granular impact is described, in part, by a velocity-dependent inertial drag.
Here the inertial drag force from the granular target is shown to be due to normal, intermittent collisions of force-carrying chains of particles with the intruder.
We confirm this collisional-based mechanism through the dependence of the drag force on intruder shape.
A collisional model for 3D impact is presented to verify our description.
The remarkable collapse of the inertial drag coefficient when accounting for intruder shape and grain density demonstrates the validity of the model.
These scaling results can lead to a complete understanding of dynamic force transmission in granular media.

We identified oscillations in the curves of inertial drag coefficient versus penetration depth that likely connect to fluctuations due to the intermittency of force chain collisions.
This was previously shown in 2D impact experiments through photoelasticity.
Direct force measurements with high sample rate are needed for the connection in 3D granular impact experiments and are the subject of future work.
Additionally, all work has focused on vertical impact.
We extend the study to oblique impact to observe the effect of a strong horizontal resistance force on penetration.

\begin{acknowledgments}
We are grateful to H. Zheng, J. Bar\'{e}s, and Y. Zhao for their assistance.
This work was supported by NASA grant no. NNX15AD38G.
C.S.B. acknowledges support from the Duke University Provost's Postdoctoral Program.

\end{acknowledgments}

\bibliography{3DImpact}

\begin{thebibliography}{27}%
\makeatletter
\providecommand \@ifxundefined [1]{%
 \@ifx{#1\undefined}
}%
\providecommand \@ifnum [1]{%
 \ifnum #1\expandafter \@firstoftwo
 \else \expandafter \@secondoftwo
 \fi
}%
\providecommand \@ifx [1]{%
 \ifx #1\expandafter \@firstoftwo
 \else \expandafter \@secondoftwo
 \fi
}%
\providecommand \natexlab [1]{#1}%
\providecommand \enquote  [1]{``#1''}%
\providecommand \bibnamefont  [1]{#1}%
\providecommand \bibfnamefont [1]{#1}%
\providecommand \citenamefont [1]{#1}%
\providecommand \href@noop [0]{\@secondoftwo}%
\providecommand \href [0]{\begingroup \@sanitize@url \@href}%
\providecommand \@href[1]{\@@startlink{#1}\@@href}%
\providecommand \@@href[1]{\endgroup#1\@@endlink}%
\providecommand \@sanitize@url [0]{\catcode `\\12\catcode `\$12\catcode
  `\&12\catcode `\#12\catcode `\^12\catcode `\_12\catcode `\%12\relax}%
\providecommand \@@startlink[1]{}%
\providecommand \@@endlink[0]{}%
\providecommand \url  [0]{\begingroup\@sanitize@url \@url }%
\providecommand \@url [1]{\endgroup\@href {#1}{\urlprefix }}%
\providecommand \urlprefix  [0]{URL }%
\providecommand \Eprint [0]{\href }%
\providecommand \doibase [0]{http://dx.doi.org/}%
\providecommand \selectlanguage [0]{\@gobble}%
\providecommand \bibinfo  [0]{\@secondoftwo}%
\providecommand \bibfield  [0]{\@secondoftwo}%
\providecommand \translation [1]{[#1]}%
\providecommand \BibitemOpen [0]{}%
\providecommand \bibitemStop [0]{}%
\providecommand \bibitemNoStop [0]{.\EOS\space}%
\providecommand \EOS [0]{\spacefactor3000\relax}%
\providecommand \BibitemShut  [1]{\csname bibitem#1\endcsname}%
\let\auto@bib@innerbib\@empty
%</preamble>
\bibitem [{\citenamefont {Poncelet}(1829)}]{poncelet}%
  \BibitemOpen
  \bibfield  {author} {\bibinfo {author} {\bibfnamefont {J.}~\bibnamefont
  {Poncelet}},\ }\href@noop {} {\emph {\bibinfo {title} {Cours de Mecanique
  Industrielle}}}\ (\bibinfo {year} {1829})\BibitemShut {NoStop}%
\bibitem [{\citenamefont {Uehara}\ \emph {et~al.}(2003)\citenamefont {Uehara},
  \citenamefont {Ambroso}, \citenamefont {Ojha},\ and\ \citenamefont
  {Durian}}]{uehara}%
  \BibitemOpen
  \bibfield  {author} {\bibinfo {author} {\bibfnamefont {J.}~\bibnamefont
  {Uehara}}, \bibinfo {author} {\bibfnamefont {M.}~\bibnamefont {Ambroso}},
  \bibinfo {author} {\bibfnamefont {R.}~\bibnamefont {Ojha}}, \ and\ \bibinfo
  {author} {\bibfnamefont {D.}~\bibnamefont {Durian}},\ }\href@noop {}
  {\bibfield  {journal} {\bibinfo  {journal} {Phys. Rev. Lett.}\ }\textbf
  {\bibinfo {volume} {90}},\ \bibinfo {pages} {194301} (\bibinfo {year}
  {2003})}\BibitemShut {NoStop}%
\bibitem [{\citenamefont {Newhall}\ and\ \citenamefont
  {Durian}(2003)}]{newhall}%
  \BibitemOpen
  \bibfield  {author} {\bibinfo {author} {\bibfnamefont {K.}~\bibnamefont
  {Newhall}}\ and\ \bibinfo {author} {\bibfnamefont {D.}~\bibnamefont
  {Durian}},\ }\href@noop {} {\bibfield  {journal} {\bibinfo  {journal} {Phys.
  Rev. E}\ }\textbf {\bibinfo {volume} {68}},\ \bibinfo {pages} {060301}
  (\bibinfo {year} {2003})}\BibitemShut {NoStop}%
\bibitem [{\citenamefont {Ambroso}\ \emph {et~al.}(2005)\citenamefont
  {Ambroso}, \citenamefont {Kamien},\ and\ \citenamefont {Durian}}]{ambroso}%
  \BibitemOpen
  \bibfield  {author} {\bibinfo {author} {\bibfnamefont {M.~A.}\ \bibnamefont
  {Ambroso}}, \bibinfo {author} {\bibfnamefont {R.~D.}\ \bibnamefont {Kamien}},
  \ and\ \bibinfo {author} {\bibfnamefont {D.~J.}\ \bibnamefont {Durian}},\
  }\href@noop {} {\bibfield  {journal} {\bibinfo  {journal} {Phys. Rev. E}\
  }\textbf {\bibinfo {volume} {72}},\ \bibinfo {pages} {041305} (\bibinfo
  {year} {2005})}\BibitemShut {NoStop}%
\bibitem [{\citenamefont {Tsimring}\ and\ \citenamefont
  {Volfson}(2005)}]{tsimring}%
  \BibitemOpen
  \bibfield  {author} {\bibinfo {author} {\bibfnamefont {L.}~\bibnamefont
  {Tsimring}}\ and\ \bibinfo {author} {\bibfnamefont {D.}~\bibnamefont
  {Volfson}},\ }\href@noop {} {\bibfield  {journal} {\bibinfo  {journal}
  {Powders and Grains}\ }\textbf {\bibinfo {volume} {2}},\ \bibinfo {pages}
  {1215} (\bibinfo {year} {2005})}\BibitemShut {NoStop}%
\bibitem [{\citenamefont {Seguin}\ \emph {et~al.}(2008)\citenamefont {Seguin},
  \citenamefont {Bertho},\ and\ \citenamefont {Gondret}}]{seguin}%
  \BibitemOpen
  \bibfield  {author} {\bibinfo {author} {\bibfnamefont {A.}~\bibnamefont
  {Seguin}}, \bibinfo {author} {\bibfnamefont {Y.}~\bibnamefont {Bertho}}, \
  and\ \bibinfo {author} {\bibfnamefont {P.}~\bibnamefont {Gondret}},\
  }\href@noop {} {\bibfield  {journal} {\bibinfo  {journal} {Phys. Rev. E}\
  }\textbf {\bibinfo {volume} {78}},\ \bibinfo {pages} {010301} (\bibinfo
  {year} {2008})}\BibitemShut {NoStop}%
\bibitem [{\citenamefont {Umbanhowar}\ and\ \citenamefont
  {Goldman}(2010)}]{umbanhowar2010}%
  \BibitemOpen
  \bibfield  {author} {\bibinfo {author} {\bibfnamefont {P.}~\bibnamefont
  {Umbanhowar}}\ and\ \bibinfo {author} {\bibfnamefont {D.}~\bibnamefont
  {Goldman}},\ }\href@noop {} {\bibfield  {journal} {\bibinfo  {journal} {Phys.
  Rev. E}\ }\textbf {\bibinfo {volume} {82}},\ \bibinfo {pages} {010301}
  (\bibinfo {year} {2010})}\BibitemShut {NoStop}%
\bibitem [{\citenamefont {Marston}\ \emph {et~al.}(2012)\citenamefont
  {Marston}, \citenamefont {Vakarelski},\ and\ \citenamefont
  {Thoroddsen}}]{thoroddsen}%
  \BibitemOpen
  \bibfield  {author} {\bibinfo {author} {\bibfnamefont {J.}~\bibnamefont
  {Marston}}, \bibinfo {author} {\bibfnamefont {I.}~\bibnamefont {Vakarelski}},
  \ and\ \bibinfo {author} {\bibfnamefont {S.}~\bibnamefont {Thoroddsen}},\
  }\href@noop {} {\bibfield  {journal} {\bibinfo  {journal} {Phys. Rev. E}\
  }\textbf {\bibinfo {volume} {86}},\ \bibinfo {pages} {020301} (\bibinfo
  {year} {2012})}\BibitemShut {NoStop}%
\bibitem [{\citenamefont {Katsuragi}\ and\ \citenamefont
  {Durian}(2007)}]{KatDur}%
  \BibitemOpen
  \bibfield  {author} {\bibinfo {author} {\bibfnamefont {H.}~\bibnamefont
  {Katsuragi}}\ and\ \bibinfo {author} {\bibfnamefont {D.}~\bibnamefont
  {Durian}},\ }\href@noop {} {\bibfield  {journal} {\bibinfo  {journal} {Nat.
  Phys.}\ }\textbf {\bibinfo {volume} {3}},\ \bibinfo {pages} {420} (\bibinfo
  {year} {2007})}\BibitemShut {NoStop}%
\bibitem [{\citenamefont {Katsuragi}\ and\ \citenamefont
  {Durian}(2013)}]{katpre2013}%
  \BibitemOpen
  \bibfield  {author} {\bibinfo {author} {\bibfnamefont {H.}~\bibnamefont
  {Katsuragi}}\ and\ \bibinfo {author} {\bibfnamefont {D.}~\bibnamefont
  {Durian}},\ }\href@noop {} {\bibfield  {journal} {\bibinfo  {journal} {Phys.
  Rev. E}\ }\textbf {\bibinfo {volume} {87}},\ \bibinfo {pages} {052208}
  (\bibinfo {year} {2013})}\BibitemShut {NoStop}%
\bibitem [{\citenamefont {Nordstrom}\ \emph {et~al.}(2014)\citenamefont
  {Nordstrom}, \citenamefont {Lim}, \citenamefont {Harrington},\ and\
  \citenamefont {Losert}}]{nordstrom}%
  \BibitemOpen
  \bibfield  {author} {\bibinfo {author} {\bibfnamefont {K.}~\bibnamefont
  {Nordstrom}}, \bibinfo {author} {\bibfnamefont {E.}~\bibnamefont {Lim}},
  \bibinfo {author} {\bibfnamefont {M.}~\bibnamefont {Harrington}}, \ and\
  \bibinfo {author} {\bibfnamefont {W.}~\bibnamefont {Losert}},\ }\href@noop {}
  {\bibfield  {journal} {\bibinfo  {journal} {Phys. Rev. Lett.}\ }\textbf
  {\bibinfo {volume} {112}},\ \bibinfo {pages} {228002} (\bibinfo {year}
  {2014})}\BibitemShut {NoStop}%
\bibitem [{\citenamefont {Altshuler}\ and\ \citenamefont {et.
  al.}(2014)}]{altshuler}%
  \BibitemOpen
  \bibfield  {author} {\bibinfo {author} {\bibfnamefont {E.}~\bibnamefont
  {Altshuler}}\ and\ \bibinfo {author} {\bibnamefont {et. al.}},\ }\href@noop
  {} {\bibfield  {journal} {\bibinfo  {journal} {Geophys. Res. Lett.}\ }\textbf
  {\bibinfo {volume} {41}},\ \bibinfo {pages} {3032} (\bibinfo {year}
  {2014})}\BibitemShut {NoStop}%
\bibitem [{\citenamefont {Tiwari}\ \emph {et~al.}(2014)\citenamefont {Tiwari},
  \citenamefont {Mohan},\ and\ \citenamefont {Sen}}]{tiwari}%
  \BibitemOpen
  \bibfield  {author} {\bibinfo {author} {\bibfnamefont {M.}~\bibnamefont
  {Tiwari}}, \bibinfo {author} {\bibfnamefont {T.~K.}\ \bibnamefont {Mohan}}, \
  and\ \bibinfo {author} {\bibfnamefont {S.}~\bibnamefont {Sen}},\ }\href@noop
  {} {\bibfield  {journal} {\bibinfo  {journal} {Phys. Rev. E}\ }\textbf
  {\bibinfo {volume} {90}},\ \bibinfo {pages} {062202} (\bibinfo {year}
  {2014})}\BibitemShut {NoStop}%
\bibitem [{\citenamefont {Li}\ \emph {et~al.}(2013)\citenamefont {Li},
  \citenamefont {Zhang},\ and\ \citenamefont {Goldman}}]{CLi}%
  \BibitemOpen
  \bibfield  {author} {\bibinfo {author} {\bibfnamefont {C.}~\bibnamefont
  {Li}}, \bibinfo {author} {\bibfnamefont {T.}~\bibnamefont {Zhang}}, \ and\
  \bibinfo {author} {\bibfnamefont {D.}~\bibnamefont {Goldman}},\ }\href@noop
  {} {\bibfield  {journal} {\bibinfo  {journal} {Science}\ }\textbf {\bibinfo
  {volume} {339}},\ \bibinfo {pages} {1408} (\bibinfo {year}
  {2013})}\BibitemShut {NoStop}%
\bibitem [{\citenamefont {Aguilar}\ and\ \citenamefont
  {Goldman}(2016)}]{aguilar}%
  \BibitemOpen
  \bibfield  {author} {\bibinfo {author} {\bibfnamefont {J.}~\bibnamefont
  {Aguilar}}\ and\ \bibinfo {author} {\bibfnamefont {D.}~\bibnamefont
  {Goldman}},\ }\href@noop {} {\bibfield  {journal} {\bibinfo  {journal}
  {Nature Physics}\ }\textbf {\bibinfo {volume} {12}},\ \bibinfo {pages} {278}
  (\bibinfo {year} {2016})}\BibitemShut {NoStop}%
\bibitem [{\citenamefont {Daniels}\ \emph {et~al.}(2004)\citenamefont
  {Daniels}, \citenamefont {Coppock},\ and\ \citenamefont
  {Behringer}}]{daniels}%
  \BibitemOpen
  \bibfield  {author} {\bibinfo {author} {\bibfnamefont {K.}~\bibnamefont
  {Daniels}}, \bibinfo {author} {\bibfnamefont {J.}~\bibnamefont {Coppock}}, \
  and\ \bibinfo {author} {\bibfnamefont {R.}~\bibnamefont {Behringer}},\
  }\href@noop {} {\bibfield  {journal} {\bibinfo  {journal} {Chaos}\ }\textbf
  {\bibinfo {volume} {14}},\ \bibinfo {pages} {S4} (\bibinfo {year}
  {2004})}\BibitemShut {NoStop}%
\bibitem [{\citenamefont {Guttler}\ \emph {et~al.}(2012)\citenamefont
  {Guttler}, \citenamefont {Hirata},\ and\ \citenamefont {Nakamura}}]{guttler}%
  \BibitemOpen
  \bibfield  {author} {\bibinfo {author} {\bibfnamefont {C.}~\bibnamefont
  {Guttler}}, \bibinfo {author} {\bibfnamefont {N.}~\bibnamefont {Hirata}}, \
  and\ \bibinfo {author} {\bibfnamefont {A.}~\bibnamefont {Nakamura}},\
  }\href@noop {} {\bibfield  {journal} {\bibinfo  {journal} {Icarus}\ }\textbf
  {\bibinfo {volume} {220}},\ \bibinfo {pages} {1040} (\bibinfo {year}
  {2012})}\BibitemShut {NoStop}%
\bibitem [{\citenamefont {Dowling}\ and\ \citenamefont
  {Dowling}(2013)}]{dowling}%
  \BibitemOpen
  \bibfield  {author} {\bibinfo {author} {\bibfnamefont {D.}~\bibnamefont
  {Dowling}}\ and\ \bibinfo {author} {\bibfnamefont {T.}~\bibnamefont
  {Dowling}},\ }\href@noop {} {\bibfield  {journal} {\bibinfo  {journal} {Am.
  J. Phys.}\ }\textbf {\bibinfo {volume} {81}},\ \bibinfo {pages} {875}
  (\bibinfo {year} {2013})}\BibitemShut {NoStop}%
\bibitem [{\citenamefont {Ruiz-Suarez}(2013)}]{ruiz}%
  \BibitemOpen
  \bibfield  {author} {\bibinfo {author} {\bibfnamefont {J.}~\bibnamefont
  {Ruiz-Suarez}},\ }\href@noop {} {\bibfield  {journal} {\bibinfo  {journal}
  {Rep. Prog. Phys.}\ }\textbf {\bibinfo {volume} {76}},\ \bibinfo {pages}
  {066601} (\bibinfo {year} {2013})}\BibitemShut {NoStop}%
\bibitem [{\citenamefont {Katsuragi}(2016)}]{KatBk}%
  \BibitemOpen
  \bibfield  {author} {\bibinfo {author} {\bibfnamefont {H.}~\bibnamefont
  {Katsuragi}},\ }\href@noop {} {\emph {\bibinfo {title} {Physics of Soft
  Impact and Cratering}}}\ (\bibinfo  {publisher} {Springer},\ \bibinfo
  {address} {Tokyo},\ \bibinfo {year} {2016})\BibitemShut {NoStop}%
\bibitem [{\citenamefont {Clark}\ \emph {et~al.}(2012)\citenamefont {Clark},
  \citenamefont {Kondic},\ and\ \citenamefont {Behringer}}]{ClarkPRL}%
  \BibitemOpen
  \bibfield  {author} {\bibinfo {author} {\bibfnamefont {A.}~\bibnamefont
  {Clark}}, \bibinfo {author} {\bibfnamefont {L.}~\bibnamefont {Kondic}}, \
  and\ \bibinfo {author} {\bibfnamefont {R.}~\bibnamefont {Behringer}},\
  }\href@noop {} {\bibfield  {journal} {\bibinfo  {journal} {Phys. Rev. Lett.}\
  }\textbf {\bibinfo {volume} {109}},\ \bibinfo {pages} {238302} (\bibinfo
  {year} {2012})}\BibitemShut {NoStop}%
\bibitem [{\citenamefont {Y.Takehara}\ \emph {et~al.}(2010)\citenamefont
  {Y.Takehara}, \citenamefont {Fujimoto},\ and\ \citenamefont
  {Okumura}}]{takehara}%
  \BibitemOpen
  \bibfield  {author} {\bibinfo {author} {\bibnamefont {Y.Takehara}}, \bibinfo
  {author} {\bibfnamefont {S.}~\bibnamefont {Fujimoto}}, \ and\ \bibinfo
  {author} {\bibfnamefont {K.}~\bibnamefont {Okumura}},\ }\href@noop {}
  {\bibfield  {journal} {\bibinfo  {journal} {EPL}\ }\textbf {\bibinfo {volume}
  {92}},\ \bibinfo {pages} {44003} (\bibinfo {year} {2010})}\BibitemShut
  {NoStop}%
\bibitem [{\citenamefont {Clark}\ \emph {et~al.}(2014)\citenamefont {Clark},
  \citenamefont {Petersen},\ and\ \citenamefont {Behringer}}]{ClarkPRE}%
  \BibitemOpen
  \bibfield  {author} {\bibinfo {author} {\bibfnamefont {A.}~\bibnamefont
  {Clark}}, \bibinfo {author} {\bibfnamefont {A.}~\bibnamefont {Petersen}}, \
  and\ \bibinfo {author} {\bibfnamefont {R.}~\bibnamefont {Behringer}},\
  }\href@noop {} {\bibfield  {journal} {\bibinfo  {journal} {Phys. Rev. E}\
  }\textbf {\bibinfo {volume} {89}},\ \bibinfo {pages} {012201} (\bibinfo
  {year} {2014})}\BibitemShut {NoStop}%
\bibitem [{\citenamefont {Nelson}\ \emph {et~al.}(2008)\citenamefont {Nelson},
  \citenamefont {Katsuragi}, \citenamefont {Mayor},\ and\ \citenamefont
  {Durian}}]{nelson}%
  \BibitemOpen
  \bibfield  {author} {\bibinfo {author} {\bibfnamefont {E.}~\bibnamefont
  {Nelson}}, \bibinfo {author} {\bibfnamefont {H.}~\bibnamefont {Katsuragi}},
  \bibinfo {author} {\bibfnamefont {P.}~\bibnamefont {Mayor}}, \ and\ \bibinfo
  {author} {\bibfnamefont {D.}~\bibnamefont {Durian}},\ }\href@noop {}
  {\bibfield  {journal} {\bibinfo  {journal} {PRL}\ }\textbf {\bibinfo {volume}
  {101}},\ \bibinfo {pages} {068001} (\bibinfo {year} {2008})}\BibitemShut
  {NoStop}%
\bibitem [{\citenamefont {Meko}(2015)}]{conv}%
  \BibitemOpen
  \bibfield  {author} {\bibinfo {author} {\bibfnamefont {D.~M.}\ \bibnamefont
  {Meko}},\ }\href@noop {} {\enquote {\bibinfo {title} {Applied time series
  analysis},}\ } (\bibinfo {year} {2015}),\ \bibinfo {note}
  {filtering}\BibitemShut {NoStop}%
\bibitem [{\citenamefont {Goldman}\ and\ \citenamefont
  {Umbanhowar}(2008)}]{goldman}%
  \BibitemOpen
  \bibfield  {author} {\bibinfo {author} {\bibfnamefont {D.}~\bibnamefont
  {Goldman}}\ and\ \bibinfo {author} {\bibfnamefont {P.}~\bibnamefont
  {Umbanhowar}},\ }\href@noop {} {\bibfield  {journal} {\bibinfo  {journal}
  {Phys. Rev. E}\ }\textbf {\bibinfo {volume} {77}},\ \bibinfo {pages} {021308}
  (\bibinfo {year} {2008})}\BibitemShut {NoStop}%
\bibitem [{\citenamefont {Clark}\ and\ \citenamefont
  {Behringer}(2013)}]{clarkepl}%
  \BibitemOpen
  \bibfield  {author} {\bibinfo {author} {\bibfnamefont {A.~H.}\ \bibnamefont
  {Clark}}\ and\ \bibinfo {author} {\bibfnamefont {R.~P.}\ \bibnamefont
  {Behringer}},\ }\href@noop {} {\bibfield  {journal} {\bibinfo  {journal}
  {EPL}\ }\textbf {\bibinfo {volume} {101}},\ \bibinfo {pages} {64001}
  (\bibinfo {year} {2013})}\BibitemShut {NoStop}%
\end{thebibliography}%

\end{document}